# Solutions of Coupled DGLAP Evolution Equations for Singlet and Gluon Structure Functions in leading order at low-x

U. Jamil[1] and J. K. Sarma[2]

Department of Physics, Tezpur University, Napam, Tezpur-784028, Assam, India

**Abstract.** In this paper, we solved the coupled Dokshitzer-Gribov-Lipatov-Altarelli-Parisi (DGLAP) evolution equations for singlet and gluon structure functions in leading order (LO) at low-x assuming the Regge behaviour of quark and gluon structure functions at this limit, and t and x-evolutions of singlet and gluon structure functions are presented. We have compared our results of gluon structure function with global MRST 2004 and GRV 1998 gluon parameterizations and the results of deuteron structure function with New Muon Collaboration (NMC) data. We have shown that the solutions of coupled equations give accurate results for the structure functions and also the compatibility of Regge behaviour of quark and gluon structure functions with perturbative quantum chromodynamics (PQCD) at low-x.

## 1 Introduction

Deep inelastic scattering (DIS) process is one of the most successful experimental methods for the understanding of quark-gluon substructure of hadrons [1-3] from which one gets the measurement of $F_2(x, Q^2)$ (proton, neutron and deuteron) structure functions in the low-x region where x is the Bjorken variable meaning the fractional momentum carried by each parton i.e., quarks and gluons, and $Q^2$ is the four momentum of the exchanged gauge boson. Moreover, structure functions are important inputs in many high energy processes and also important for examination of PQCD [3], the underlying dynamics of quarks and gluons. In PQCD, for high-$Q^2$, the $Q^2$-evolutions of these densities (at fixed-x) are given by the DGLAP evolution equations [4, 5]. These equations introduced the parton distribution functions which can be interpreted as the probability of finding, say in a proton, respectively a quark, an antiquark or a gluon with four momentum $Q^2$ and momentum fraction x, and these are often considered as a very good test of PQCD. The structure function $F_2(x, Q^2)$ reflects the momentum distribution of the quarks in the nucleon, an important aspect of its internal structure. Measurement of the structure function as a function of x and $Q^2$ yields information on the shape of the parton distribution functions. The solutions of the DGLAP equations can be calculated either by numerical integration in steps or by taking the moments of the distributions [6]. Among various solutions of this equation, most of the methods are numerical. Mellin moment space [7] with subsequent inversion, Brute force method [8], Laguerre method [9], Matrix method [10] etc. are different methods used to solve DGLAP evolution equations. The short comings common to all are the computer time required and decreasing accuracy for x→0 [10]. More precise approach is the matrix approach to the solution of the DGLAP evolution equations, yet it is also a numerical solution. Thus though numerical solutions are available in the literature, the explorations of the possibilities of obtaining analytical solutions of DGLAP evolution equations are always interesting. Some approximated analytical solutions of DGLAP evolution equations suitable at low-x, have been reported in recent years [11, 12] with considerable phenomenological success. And structure functions thus calculated are expected to rise approximately with a power of x towards low-x which is supported by Regge theory [13, 14]. The low-x region of DIS offers a unique possibility to explore the Regge limit [13] of PQCD. The low-x behaviour (at fixed-$Q^2$) of parton distributions can be considered by a triple pole pomeron model [14, 15] at the initial scale $Q_0^2$ and then evolved using DGLAP equations. The Regge behaviour of the sea quark and antiquark distributions is given by $q_{sea}(x) \sim x^{\lambda p}$ with pomeron exchange [14] of intercept $\lambda p = -1$. But the valence quark distribution for low-x given by $q_{val}(x) \sim x^{-\lambda r}$ corresponding to a reggeon exchange of intercept $\lambda r = 1/2$.

---

[1]jamil@tezu.ernet.in, [2]jks@tezu.ernet.in

In the analytical solutions of DGLAP evolution equations for singlet or gluon structure functions, a relation between singlet structure function and gluon structure function has to be assumed [11,12,16,17]. The commonly used relation is $F_2^S(x,t)=K(x) G(x,t)$, where K(x) is an ad hoc function of x. Since these evolution equations of gluon and singlet structure functions in leading order are in the same forms of derivative with respect to t, so we can consider this function. And also the input singlet and gluon parameterizations, taken for the global analysis to incorporate different high precision data, are also functions of x at fixed $Q^2$ [18, 19]. So the relation between singlet and gluon structure functions will come out in terms of x at fixed-$Q^2$. So to get the solution we can assume some simple standard functional forms of K(x), yet we do not know the actual form. The actual functional form of K(x) can be determined by simultaneous solutions of coupled equations of singlet and gluon structure functions. So to overcome the assumption of this ad hoc function K(x), in our present work, we have derived the solution of coupled DGLAP evolution equations for singlet and gluon structure functions at low-x in leading order (LO) considering Regge behaviour of structure functions. The t and x-evolutions of singlet and gluon structure functions thus obtained have been compared with NMC deuteron data and global MRST 2004 and GRV 1998 gluon parameterizations respectively. Here we have shown that the solutions of coupled equations give almost accurate results for the structure functions. In this paper, section 1, section 2, section 3 and section 4 are the introduction, theory, results and discussion, and conclusions respectively.

## 2 Theory

The LO DGLAP evolution equations for singlet and gluon structure functions have the standard forms [20, 21] respectively

$$Q^2 \frac{\partial}{\partial Q^2} F_2^S(x, Q^2) = \frac{\alpha_s(Q^2)}{3\pi} \Big\{ (3 + 4\ln(1-x)) F_2^S(x, Q^2)$$

$$+ \int_x^1 d\omega \Big[ \frac{2}{(1-\omega)} \{(1+\omega^2) F_2^S(x/\omega, Q^2) - 2 F_2^S(x, Q^2)\} + \frac{3}{2} N_f \int_x^1 \{\omega^2 + (1-\omega)^2\} G(x/\omega, Q^2) d\omega \Big] \Big\} \quad (1)$$

and

$$Q^2 \frac{\partial}{\partial Q^2} G(x, Q^2) = \frac{3\alpha_s(Q^2)}{\pi} \Big\{ \Big( \frac{11}{12} - \frac{N_f}{18} + \ln(1-x) \Big) G(x, Q^2)$$

$$+ \int_x^1 d\omega \Big[ \frac{\omega G(x/\omega, Q^2) - G(x, Q^2)}{1-\omega} + \Big( \omega(1-\omega) + \frac{1-\omega}{\omega} \Big) G(x/\omega, Q^2) + \frac{2}{9} \Big( \frac{1+(1-\omega)^2}{\omega} \Big) F_2^S(x/\omega, Q^2) \Big] \Big\}, \quad (2)$$

where $\alpha_s(Q^2) = \frac{12\pi}{(33 - 2N_f) \ln(Q^2/\Lambda^2)}$, $\Lambda$ being the QCD cut-off parameter and $N_f$ is the number of flavours. After changing the variable $Q^2$ by t where $t = \ln(Q^2/\Lambda^2)$, we get

$$\frac{\partial F_2^S(x, t)}{\partial t} - \frac{A_f}{t} \Big[ \{3 + 4\ln(1-x)\} F_2^S(x, t) + 2 \int_x^1 \frac{d\omega}{1-\omega} \{(1+\omega^2) F_2^S(x/\omega, t) - 2 F_2^S(x, t)\}$$

$$+ \frac{3}{2} N_f \int_x^1 \{\omega^2 + (1-\omega)^2\} G(x/\omega, t) d\omega \Big] = 0 \quad (3)$$

and

$$\frac{\partial G(x, t)}{\partial t} - \frac{9A_f}{t}\left\{\left(\frac{11}{12} - \frac{N_f}{18} + \ln(1-x)\right)G(x, t) + I_g\right\} = 0, \tag{4}$$

where

$$I_g = \int_x^1 d\omega\left[\frac{\omega G(x/\omega, t) - G(x,t)}{1-\omega} + \left(\omega(1-\omega) + \frac{1-\omega}{\omega}\right)G(x/\omega, t) + \frac{2}{9}\left(\frac{1+(1-\omega)^2}{\omega}\right)F_2^S(x/\omega, t)\right],$$

and $A_f = 4/(33 - 2N_f)$.

Now let us consider the Regge behaviour of singlet and gluon structure functions [14, 22, 23] as

$$F_2^S(x, t) = T_1(t) x^{-\lambda_S} \tag{5}$$

and

$$G(x, t) = T_2(t) x^{-\lambda_G}, \tag{6}$$

where $T_1(t)$ and $T_2(t)$ are functions of t, and $\lambda_S$ and $\lambda_G$ are the Regge intercepts for singlet and gluon structure functions respectively. From equations (5) and (6) we get

$$F_2^S(x/\omega, t) = T_1(t) \omega^{\lambda_S} x^{-\lambda_S} \tag{7}$$

and

$$G(x/\omega, t) = T_2(t) \omega^{\lambda_G} x^{-\lambda_G}. \tag{8}$$

Putting equations (5) to (8) in equations (3) and (4), we get respectively

$$\frac{\partial T_1(t)}{\partial t} = \frac{T_1(t)}{t} f_1(x) + \frac{T_2(t)}{t} f_2(x) \tag{9}$$

and

$$\frac{\partial T_2(t)}{\partial t} = \frac{T_1(t)}{t} f_3(x) + \frac{T_2(t)}{t} f_4(x), \tag{10}$$

where

$$f_1(x) = A_f\left[\{3 + 4\ln(1-x)\} + 2\int_x^1 \frac{d\omega}{1-\omega}\left\{(1+\omega^2)\omega^{\lambda_S} - 2\right\}\right],$$

$$f_2(x) = x^{(\lambda_S - \lambda_G)}\left[\frac{3}{2} \cdot A_f \cdot N_f \int_x^1 \{\omega^2 + (1-\omega)^2\} \cdot \omega^{\lambda_G} d\omega\right],$$

$$f_3(x) = 9A_f\left\{(A + \ln(1-x)) + \int_x^1 d\omega\left[\frac{(\omega^{\lambda_G + 1} - 1)}{1-\omega} + \left(\omega(1-\omega) + \frac{1-\omega}{\omega}\right)\omega^{\lambda_G}\right]\right\}$$

and

$$f_4(x) = 9A_f \cdot \frac{2}{9} \cdot x^{(\lambda_G - \lambda_S)} \cdot \int_x^1 \left( \frac{1 + (1-\omega)^2}{\omega} \right) \omega^{\lambda_S} d\omega \ .$$

Let us take, $f_1(x)=P$, $f_2(x)=Q$, $f_3(x)=R$, $f_4(x)=S$. Equations (9) and (10) have got the simple forms respectively as

$$t \cdot \frac{\partial T_1(t)}{\partial t} - P \cdot T_1(t) - Q \cdot T_2(t) = 0 \tag{11}$$

and

$$t \cdot \frac{\partial T_2(t)}{\partial t} - R \cdot T_1(t) - S \cdot T_2(t) = 0. \tag{12}$$

Equations (11) and (12) are simultaneous linear ordinary differential equations in $T_1(t)$ and $T_2(t)$. Solving these equations by one of the standard methods for solution of ordinary differential equations [24, 25], we arrived at

$$T_1(t) = C_1 t^{g_1} + C_2 t^{g_2} \tag{13}$$

$$T_2(t) = C_1 F_1 t^{g_1} + C_2 F_2 t^{g_2}, \tag{14}$$

where $C_1$ and $C_2$ are arbitrary constants, $g_1 = \frac{-(u-1) + \sqrt{(u-1)^2 - 4v}}{2}$ and $g_2 = \frac{-(u-1) - \sqrt{(u-1)^2 - 4v}}{2}$, where u = 1-P-S, v = S.P-Q.R, $F_1 = (g_1-P)/Q$, $F_2 = (g_2-P)/Q$. As $C_1$ and $C_2$ are only arbitrary constants, we can take $C_1 = C_2 = C$. Hence the forms of singlet and gluon structure functions become

$$F_2^S(x,t) = C(t^{g_1} + t^{g_2}) x^{-\lambda_S} \tag{15}$$

and

$$G(x,t) = C(F_1 t^{g_1} + F_2 t^{g_2}) x^{-\lambda_G}. \tag{16}$$

Applying initial conditions at $x = x_0$, $F_2^S(x,t) = F_2^S(x_0, t)$ and $G(x,t) = G(x_0, t)$, and at $t = t_0$, $F_2^S(x,t) = F_2^S(x, t_0)$ and $G(x,t) = G(x, t_0)$, we found the t and x-evolution equations for the singlet and gluon structure functions respectively as

$$F_2^S(x,t) = F_2^S(x, t_0) \frac{(t^{g_1} + t^{g_2})}{(t_0^{g_1} + t_0^{g_2})} \ , \tag{17}$$

$$F_2^S(x,t) = F_2^S(x_0,t) \frac{\left(t^{g_1} + t^{g_2}\right)}{\left(t^{g_{10}} + t^{g_{20}}\right)} \left(\frac{x_0}{x}\right)^{\lambda_S}, \tag{18}$$

$$G(x,t) = G(x,t_0) \frac{\left(F_1 t^{g_1} + F_2 t^{g_2}\right)}{\left(F_1 t_0^{g_1} + F_2 t_0^{g_2}\right)} \tag{19}$$

and

$$G(x,t) = G(x_0,t) \frac{\left(F_1 t^{g_1} + F_2 t^{g_2}\right)}{\left(F_{10} t^{g_{10}} + F_{20} t^{g_{20}}\right)} \left(\frac{x_0}{x}\right)^{\lambda_G}, \tag{20}$$

where $g_{10}$, $g_{20}$, $F_{10}$ and $F_{20}$ are the values of $g_1$, $g_2$, $F_1$ and $F_2$ at $x = x_0$. The deuteron $F_2$ structure functions measured in DIS can be written in terms of singlet structure functions as $F_2^d = (5/9) F_2^S$ [21]. Hence, the t and x-evolution equations of deuteron structure function are respectively

$$F_2^d(x,t) = F_2^d(x,t_0) \frac{\left(t^{g_1} + t^{g_2}\right)}{\left(t_0^{g_1} + t_0^{g_2}\right)}, \tag{21}$$

and

$$F_2^d(x,t) = F_2^d(x_0,t) \frac{\left(t^{g_1} + t^{g_2}\right)}{\left(t^{g_{10}} + t^{g_{20}}\right)} \left(\frac{x_0}{x}\right)^{\lambda_S}. \tag{22}$$

## 3 Results and discussion

This paper presents the t and x-evolutions of singlet and gluon structure functions at low-x, which are obtained by solving coupled DGLAP equations applying Regge behaviour of structure functions given by equations (17) to (20) respectively. We compare our results of t and x-evolutions of gluon distribution function in LO given by equations (19) and (20) respectively with MRST 2004 and GRV 1998 parameterizations. We have taken the MRST 2004 fit [26] to the ZEUS [27] and H1 [28] data with x<0.01 and $2<Q^2<500$ GeV$^2$ for $Q^2 =$ 100 GeV$^2$, in which they have taken the MRST-like parametric form same as for MRST 2001 fit [18] for the starting distribution at $Q_0^2 = 1$ GeV$^2$ given by xg = $A_g x^{-\lambda_g}(1-x)^{3.7}(1+\varepsilon_g\sqrt{x}+\gamma_g x) - A x^{-\delta}(1-x)^{10}$, the optimum fit corresponds to $\alpha_s(M_z^2) = 0.119$ i.e. $\Lambda_{\overline{MS}}$ ($N_f$ = 4) = 323 MeV. The $\lambda_g$, $\varepsilon_g$, A and $\delta$ are taken as free parameters with $A_g$ determined from the momentum sum rule and $\gamma_g$ initially fixed at zero and they have taken the MRST-like inputs $A_g$=10.1, $\lambda_g$=(-0.49±0.1), $\varepsilon_g$=(-1.2±0.1), A=(2.4±5.8)×10$^{-3}$ and $\delta$=(0.74±0.3). We have taken the GRV 1998 parameterization [19] for $10^{-2} \leq x \leq 10^{-5}$ and $20 \leq Q^2 \leq 80$ GeV$^2$, where they used H1 [29] and ZEUS [30] high precision data on G(x, Q$^2$). They have chosen $\alpha_s(M_z^2)$ = 0.114 i.e. $\Lambda_{\overline{MS}}$ (N$_f$ = 4) = 246 MeV. The input densities have been fixed using the data sets of HERA [29], SLAC [31], BCDMS [32], NMC [33] and E665 [34]. The resulting input distribution at Q$^2$ = 0.40 GeV$^2$ is given by xg = 20.80x$^{1.6}$(1–x)$^{4.1}$. We compare our results of t and x-evolutions of deuteron structure function in LO from equation (21) and (22) with the NMC small-x medium-Q$^2$ data [33] respectively. Deuteron structure function $F_2^d(x, Q^2)$ measured in the range of 15$\leq$Q$^2$$\leq$27 GeV$^2$, 0.0175$\leq$x$\leq$0.0045 have been used for phenomenological analysis of deuteron structure functions. Here we used the QCD cut-off parameter $\Lambda_{\overline{MS}}$(N$_f$ = 4) = 323 MeV for $\alpha_s(M_z^2)$ = 0.119± 0.002 [18]. According to Regge theory, the high energy (low-x) behaviour of both gluons and sea quarks is controlled by the same singularity factor in the complex angular momentum plane [14, 26], and so we would expect $\lambda_S = \lambda_G = \lambda$. And as the value of $\lambda$ should be close to 0.5 in quite a broad range of low-x [14, 22], we would also expect that our theoretical curves are best fitted to those of the experimental data and parameterization curves at $\lambda_S = \lambda_G = \lambda \approx 0.5$.

In figures 1(a) and 1(b), we compare our result of t-evolution of gluon distribution function from equation (19) with GRV 1998 gluon distribution parameterization at x=10$^{-5}$ and 10$^{-4}$ respectively. The best fit results correspond to $\lambda_S = \lambda_G = 0.47$ for x=10$^{-5}$ and $\lambda_S = \lambda_G = 0.5$ for x=10$^{-4}$. The figures show good agreement of our result with GRV 1998 parameterization at low-x.

In figures 2(a) we compare our result of x-evolution of gluon structure function from equation (20) with MRST 2004 gluon distribution parameterizations at Q$^2$ = 100 GeV$^2$. We find the best fit result corresponding to $\lambda_S$ =0.2, $\lambda_G$ = 0.7. Here as x-values are moderately low, so the fitting is not so good. In figures 2(b) to 2(d), we compare our result of x-evolution of gluon structure function from equation (20) with GRV 1998 gluon distribution parameterization at Q$^2$ = 20 GeV$^2$, 40 GeV$^2$ and 80 GeV$^2$ respectively. For all the three figures the best fit results are for $\lambda_S$ =0.44, $\lambda_G$ = 0.64. From the figures it is obvious that our result is best fitted to the GRV 1998 parameterizations for increasing Q$^2$ at low-x.

In some recent papers [35], Choudhury and Saharia presented a form of gluon distribution function at low-x obtained from a unique solution with one single initial condition through the application of the method of characteristics [36]. They have overcome the limitations of non-uniqueness of some of the earlier approaches [11, 16]. So, it is theoretically and phenomenologically favoured over the earlier approximations. We have presented this result with GRV 1998 parameterization and our result, and found that with decreasing x we get a better fit of our result to GRV 1998 parameterization in comparison with their result.

In figures 3(a) and 3(b), we compare our result for t-evolution of deuteron structure function in LO from equation (21) with NMC data. The best fit result corresponds to $\lambda_S = \lambda_G = 0.5$ for x=0.0045 and $\lambda_S = 0.5$, $\lambda_G = 0.55$ for x=0.0175. In figures 3(c) and 3(d), we compare our results for x-evolution of deuteron structure function in LO from equation (22) with NMC data. We find the best fit result correspondings to $\lambda_S = 0.1$, $\lambda_G = 0.9$ for $Q^2 = 20$ GeV$^2$ and $\lambda_S = 0.2$, $\lambda_G = 0.8$ for $Q^2 = 27$ GeV$^2$. As Regge theory strictly applicable only for low-x and high-$Q^2$ [14, 37], the best fits of our result for x-evolution of deuteron structure function with NMC data are not so good and λ values are also not close to 0.5. As though $Q^2$ values are moderately high, but corresponding x values are also high. Due to lack of deuteron data at low-x and high-$Q^2$, we could not check our result for x-evolution of deuteron structure function properly. But from the two best fit graphs presented here we can see that our result approaches NMC data for higher $Q^2$ at low-x.

Figures 4(a) and 4(b) show the sensitivity of the parameters $\lambda_S$ and $\lambda_G$ respectively. Taking the best fit figures to the t-evolution of gluon distribution function of GRV 1998 parameterization for x=$10^{-4}$, we have given the results for the ranges of the parameters as $0.35 \leq \lambda_S \leq 0.65$ and $0.45 \leq \lambda_G \leq 0.55$. It is observed that $\lambda_G$ is more sensitive than $\lambda_S$.

**4 Conclusions**

Here we have obtained a new description of t and x-evolutions of both the singlet and gluon structure functions solving the coupled evolution equations within the Regge limit. We have seen that our results are in good agreement with NMC data and MRST 2004 and GRV 1998 global parameterizations especially at low-x and high-$Q^2$ region. We can conclude that Regge behaviour of quark and gluon distribution functions is compatible with PQCD at that region assuming the Regge intercept almost same for both quark and gluon. Considering Regge behaviour of structure functions DGLAP equations become quite simple to solve and also the solution of coupled DGLAP evolution equations becomes possible. So this method is a viable simple alternative to other methods. Here we overcome the problem of ad hoc assumption of the function K(x). From the expressions of structure functions thus obtained (given by equations (15) and (16)) the exact form of K(x) may be worked out. Moreover, here we solve only leading order evolution equations. We expect that next-to-leading order equations are more correct and their solutions will give better fit to global data and parameterizations. Again as the form of evolution equations for spin dependent structure functions is also same as for spin independent structure functions, we hope this method will be applicable to spin dependent cases also.

24. J.N. Sharma and Dr. R.K. Gupta, 'Differential Equations',Krishna Prakashan Mandir, Meerut, 1990
25. I. Sneddon, 'Elements of Partial Differential Equations', McGraw-Hill, New York, 1957
26. A. D. Martin, M. G. Ryskin and G. Watt, arXiv:hep-ph/0406225 (2004)
27. S. Chekanov et al. [ZEUS Collaboration], Eur. Phys. J. C21, 443 (2001)
28. C. Adloff et al. [H1 Collaboration], Eur. Phys. J. C21, 33 (2001)
29. S. Aid et al., H1 Collaboration, Nucl. Phys. B470, 3 (1996)
30. M. Derrick et. al., ZEUS Collaboration, Z. Phys. C69, 607 (1996)
31. L. W. Whitlow et al., Phys. Lett. B282, 475 (1992) ; SLAC- report 357 (1990)
32. A. C. Benvenuti et al., BCDMS Collaboration, Phys. Lett. B223, 485 (1989)
33. M. Arneodo et al., NMC Collaboration, Nucl. Phys. B483, 3 (1997)
34. M. R. Adams et al., E665 Collaboration, Phys. Rev. D54, 1006(1996)
35. D. K. Choudhury and P. K. Sahariah, Pramana-J. Phys. 58, 599 (2002); 60, 563 (2003)
36. S. J. Farlow, 'Partial differential equations for scientists and engineers' (John Willey, New York 1982)
37. G. Soyez, Phys. Rev. D69, 096005 (2004); hep-ph/0306113 (2003)
**Figure captions**

**Fig. 1.** t-evolution of gluon distribution function in LO at $x=10^{-5}$ and $10^{-4}$. Data points at lowest-$Q^2$ values are taken as input to test the evolution equation (19). Here Fig. 1(a) and 1(b) are the best fit graphs of our result with GRV 1998 parameterization at $x=10^{-5}$ and $10^{-4}$ respectively.

**Fig. 2 .** x-evolution of gluon structure function in LO. Data points for $x \leq 0.1$ are taken as inputs to test the evolution equation (20). Here Fig. 2(a) is the best fit graph of our result with MRST 2004 parameterization for $Q^2 = 100$ GeV$^2$. Fig. 2(b) to 2(d) are the best fit graphs of our result with GRV 1998 parameterization for $Q^2 = 20, 40$ and $80$ GeV$^2$ respectively.

**Fig. 3.** t and x-evolutions of deuteron structure function in LO for the representative values of $Q^2$ and x respectively. Data points at lowest-$Q^2$ values are taken as input to test the evolution equation (21) and data points for $x \leq 0.1$ are taken as input to test the evolution equation (22). Here Fig. 3(a) to 3(d) are the best fit graphs of our results with NMC data.

**Fig. 4.** Fig. 4(a) to 4(b) show the sensitivity of the parameters $\lambda_S$ and $\lambda_G$ at $Q^2= 80$ GeV$^2$ and $x=10^{-4}$ respectively.

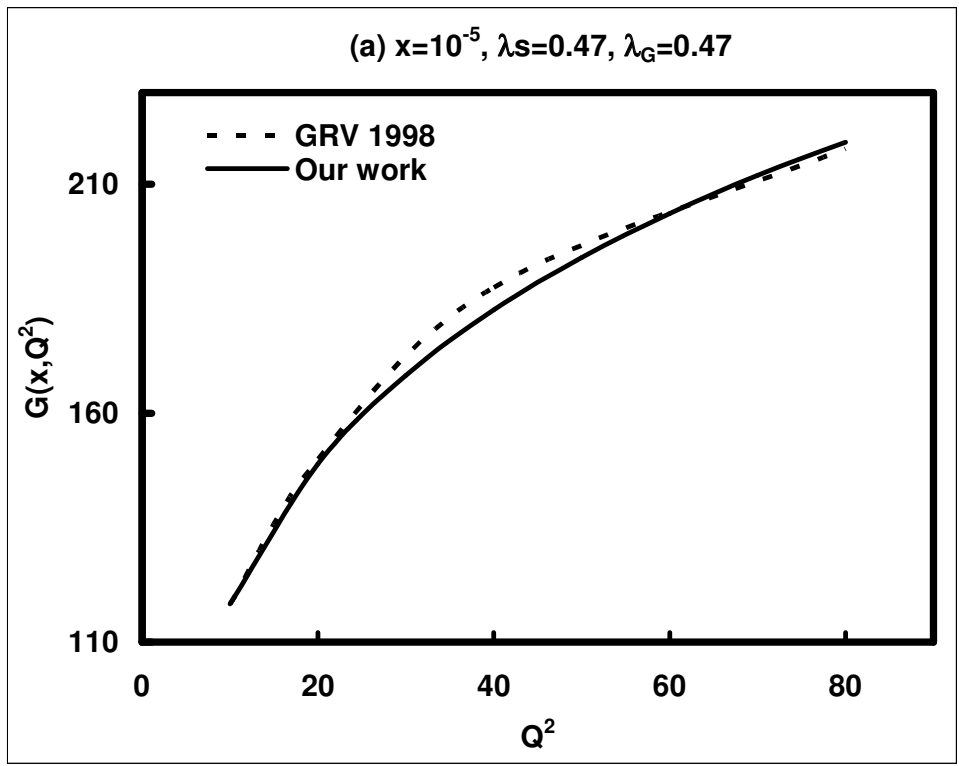

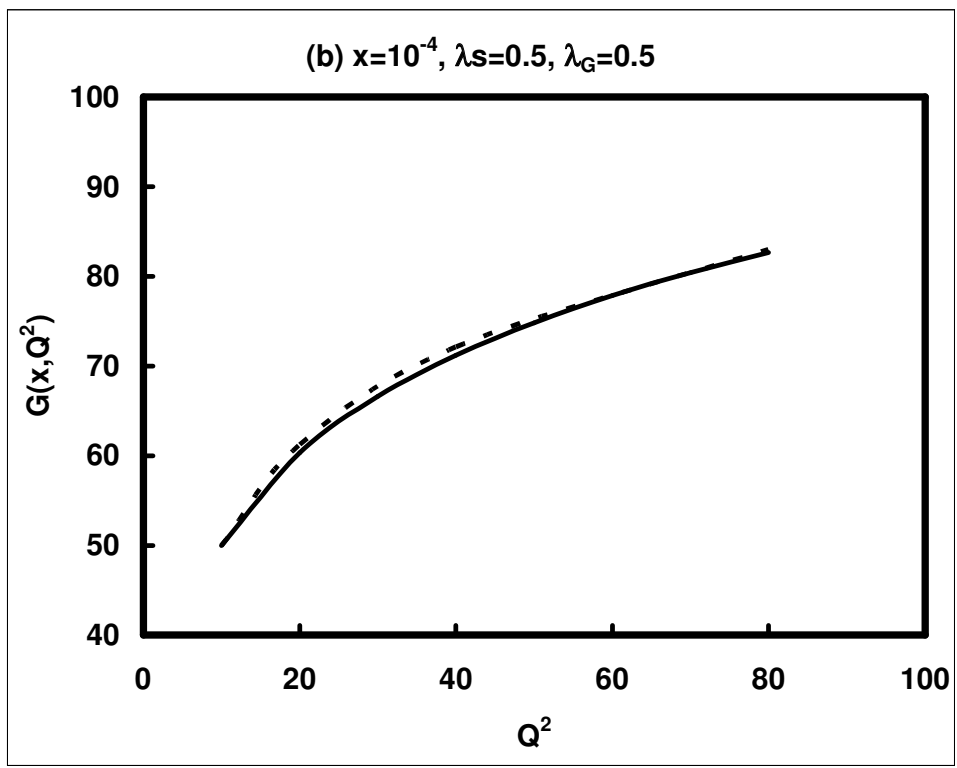

**Fig.1**

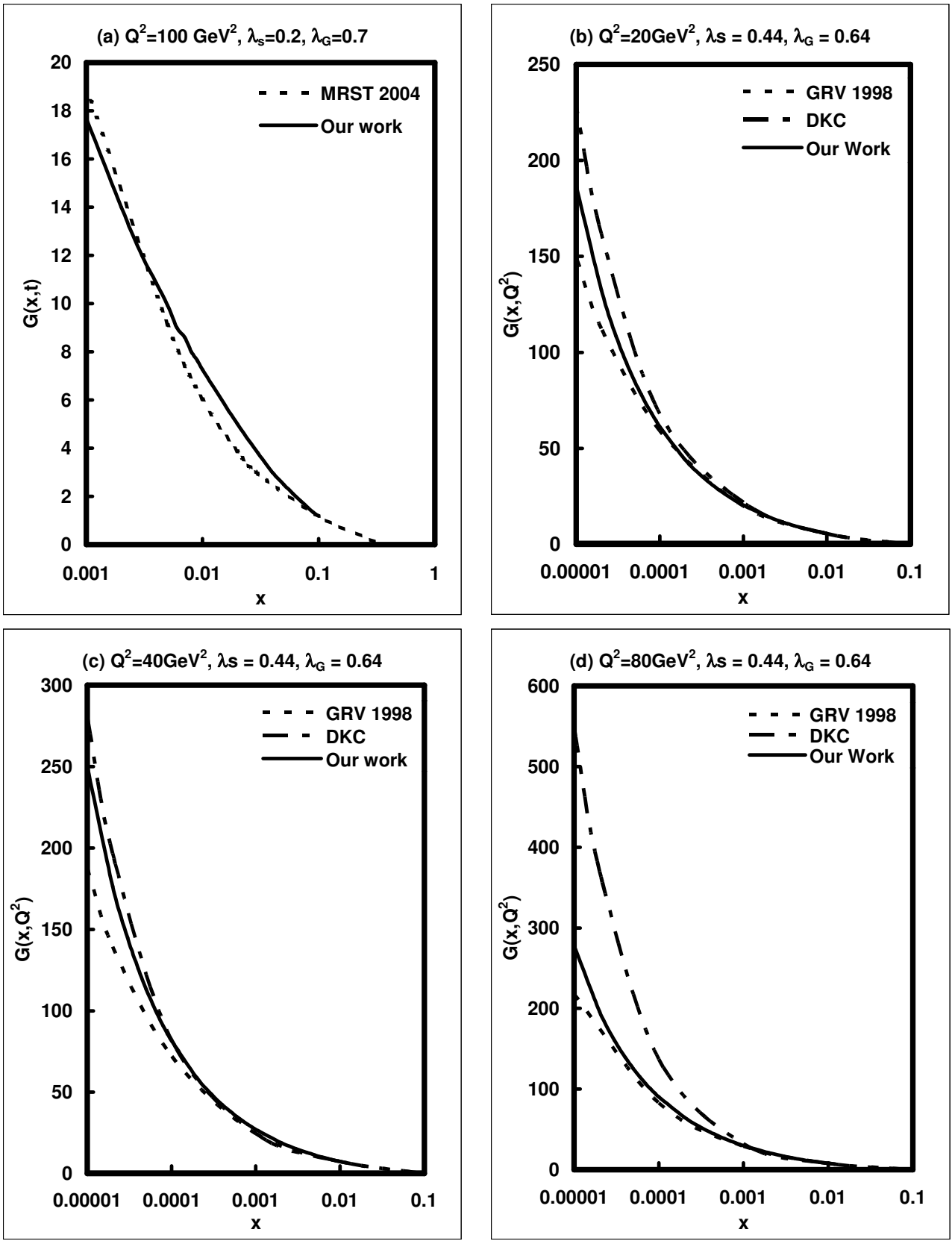

**Fig.2**

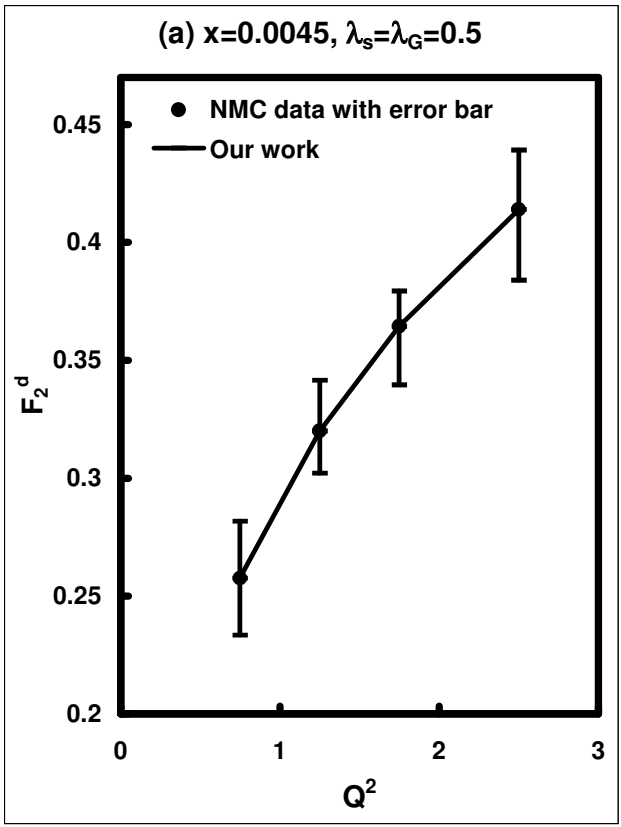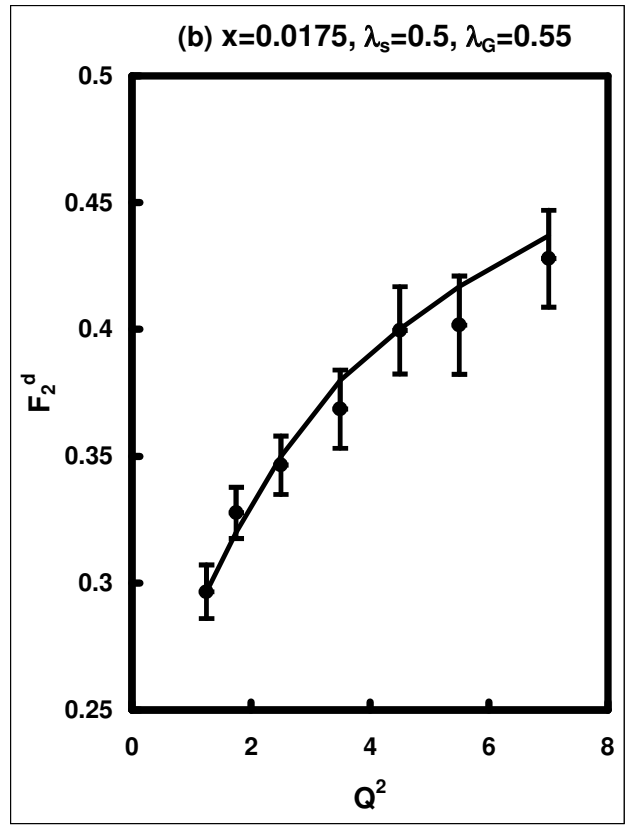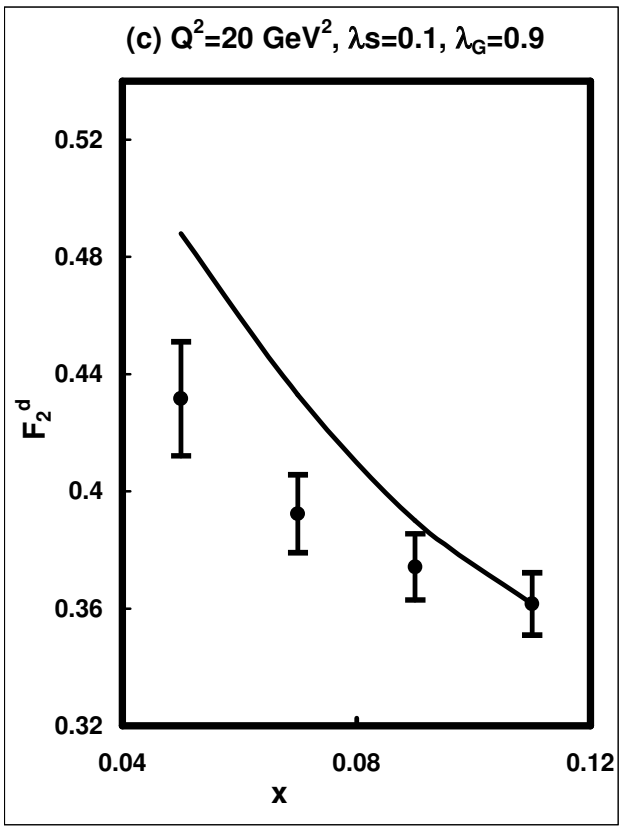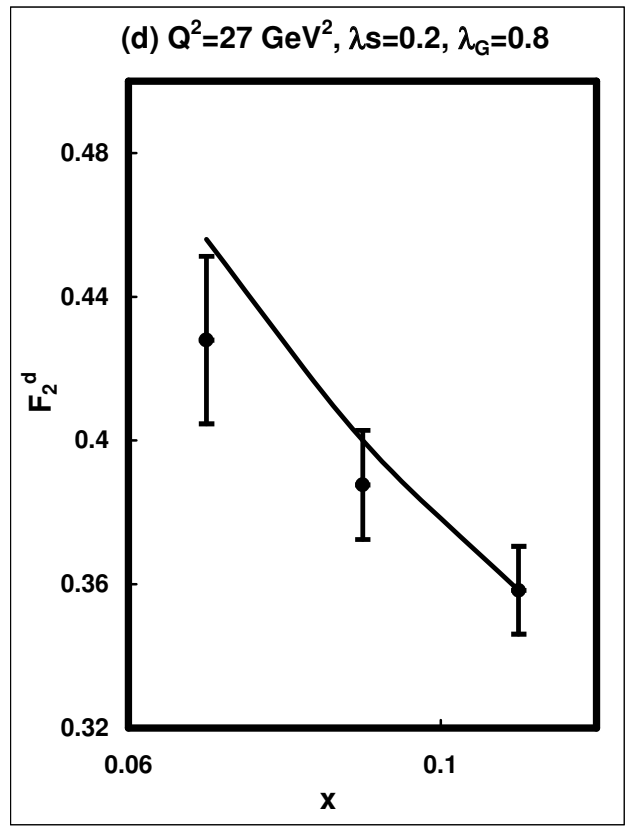

**Fig.3**

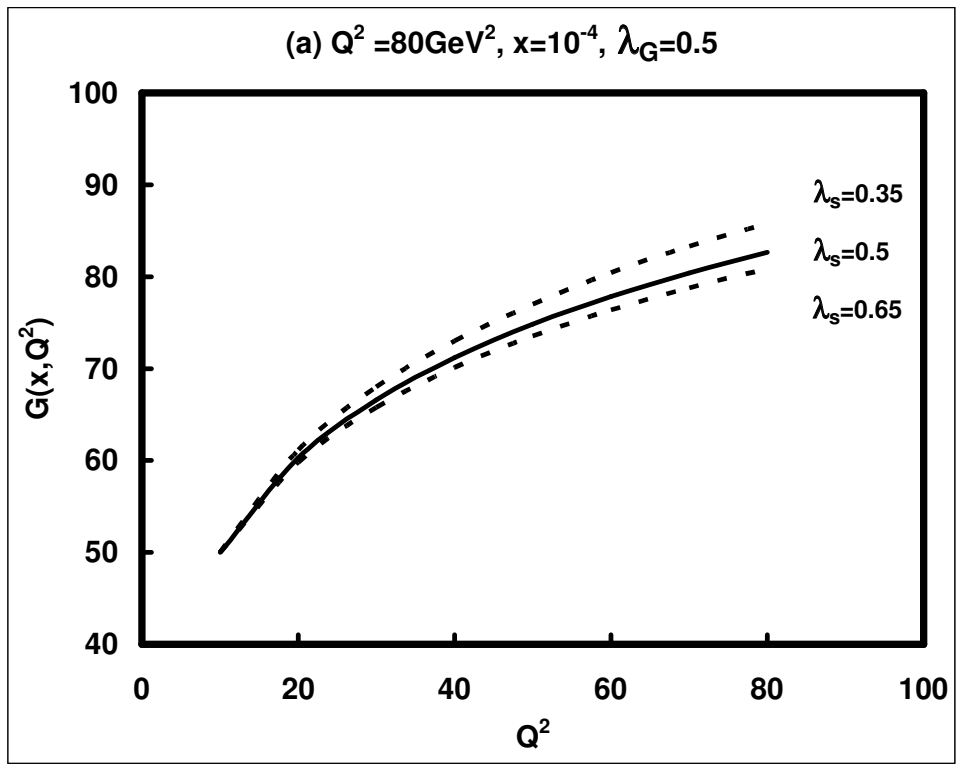
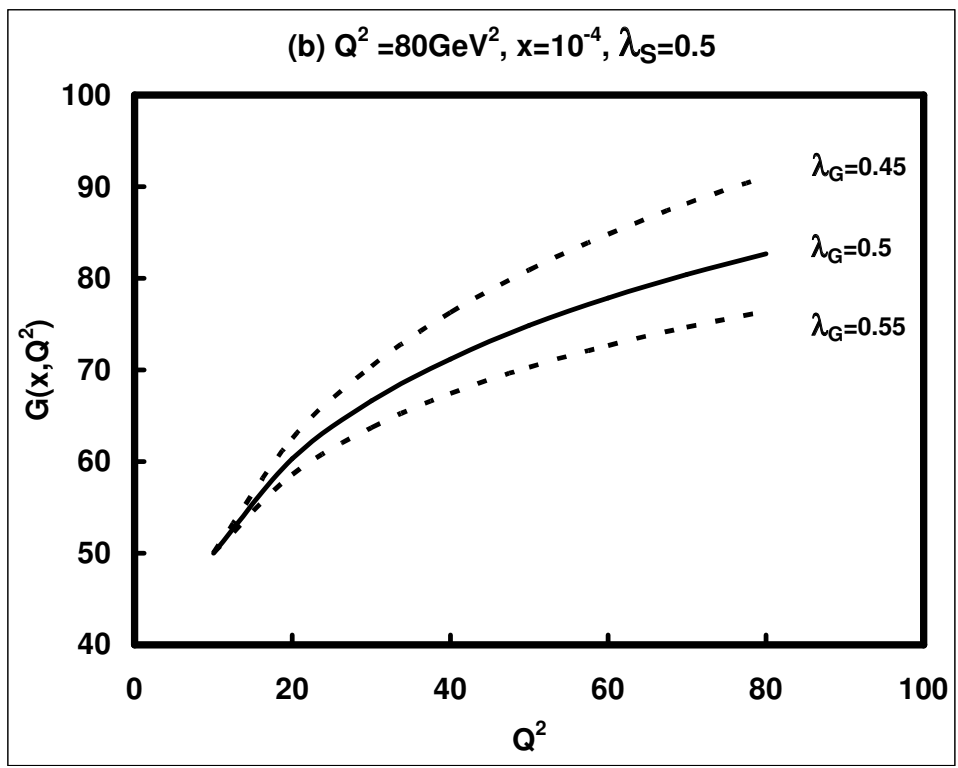

**Fig.4**